\documentclass[aps,prb,reprint,superscriptaddress,longbibliography]{revtex4-2}

\usepackage{amsmath}
\usepackage{amssymb}
\usepackage{graphicx}
\usepackage{bm}
\usepackage{microtype}
\newenvironment{manuscriptfigure}{\begin{figure*}[t]\centering}{\end{figure*}}
\newenvironment{manuscripttable}{\begin{table}[tb]}{\end{table}}

\begin{document}
\raggedbottom
\clubpenalty=10000
\widowpenalty=10000
\displaywidowpenalty=10000
\setlength{\parskip}{0pt}
\setlength{\dbltextfloatsep}{10pt plus 2pt minus 2pt}
\setlength{\abovecaptionskip}{6pt}

\title{Core Reactivity in Cs-Ba-O ternary system under pressure}

\author{Samantha Ellis}
\affiliation{Department of Chemistry and Biochemistry, California State University, Northridge, California 91330, USA}

\author{Austin Ellis}
\affiliation{Department of Chemistry and Biochemistry, California State University, Northridge, California 91330, USA}

\author{Lansen Wyke}
\affiliation{Department of Chemistry and Biochemistry, California State University, Northridge, California 91330, USA}

\author{Christina Taylor}
\affiliation{Department of Chemistry and Biochemistry, California State University, Northridge, California 91330, USA}

\author{Maosheng Miao}
\email[Corresponding author: ]{mmiao@csun.edu}
\affiliation{Department of Chemistry and Biochemistry, California State University, Northridge, California 91330, USA}

\date{\today}

\begin{abstract}
High pressure can activate Cs $5p$ core electrons for covalent bonding. In this study, we investigate the Cs--Ba--O system at 150 and 250~GPa using first-principles structure searches. CsBa$_3$O$_8$ and CsBaO$_4$ both lie on the static-lattice convex hull at 250~GPa and exhibit mild Cs $5p$--O $2p$ bonding while retaining O$_2$ units with intermediate superoxide--peroxide character. Cs$_2$BaO$_3$ shows substantially stronger bonding but remains metastable with respect to CsO and BaO. A 1:1 Cs$_2$O--BaO$_2$ mixture provides a solid starting composition for investigating this chemistry, with core-reactive CsO among the predicted products at 250~GPa. Calculated Cs $L_3$-edge x-ray absorption near-edge spectra and powder diffraction patterns provide electronic and structural signatures to guide identification of the predicted phases in diamond anvil cell experiments.
\end{abstract}

\maketitle

\section{Introduction}

High pressure can significantly alter the chemical behavior of elements. By shifting orbital energies and redistributing electrons, compression can favor unconventional compositions and bonding~\cite{Rahm2019Squeezing,Zhang2013SodiumChlorides,Flosbach2025Pr2O5,Zurek2009LithiumHydrogen}, induce electride behavior, and reverse conventional redox roles~\cite{Ma2009TransparentSodium,Storm2025SodiumElectride,Ellis2026HydrogenElectride,Wang2025FeRedox}.

Pressure can activate inner-shell electrons for bonding~\cite{Miao2013CsFluorides,Racioppi2025Semicore}. Cs has spatially extended $5p$ orbitals that can participate in covalent bonding under compression, as predicted for Cs fluorides~\cite{Miao2013CsFluorides}.

In cesium oxides, CsO is predicted to transform near 221~GPa from a peroxide structure into an $R\overline{3}m$ phase with nearly linear O--Cs--O units and substantial Cs $5p$--O $2p$ bonding. The O--O dimers disappear while the composition remains unchanged. Compression of solid CsO therefore offers a route to core reactivity without loading the highly corrosive molecular fluorine required for the original Cs-fluoride proposal~\cite{Sun2021CsPolyoxides}.

As an alternative to preformed CsO, we consider Cs$_2$O and BaO$_2$ as solid starting materials. BaO$_2$ supplies oxygen as a solid peroxide, motivating the choice of Ba as the third element. It is commercially available, while Cs$_2$O has established preparative routes under inert conditions~\cite{Gemming2005Cs2O}. Assessing a proposed oxide mixture requires comparing formation of the desired compound with formation of competing products~\cite{McDermott2023Selectivity}. A 1:1 mixture of Cs$_2$O and BaO$_2$ corresponds to Cs$_2$BaO$_3$, but the same overall composition can also form CsO and BaO. The ternary phase diagram identifies the compounds favored at this composition, while their electronic structures reveal whether Cs core participation occurs in the ternary or a competing binary phase.

Here, we examine the Cs--Ba--O system using crystal structure searches and ternary convex hulls at 150 and 250~GPa. We identify thermodynamically stable ternaries with weaker Cs $5p$ participation and a metastable Cs$_2$BaO$_3$ structure with substantially stronger, site-dependent Cs $5p$--O $2p$ bonding. 

We compare the ternaries with three binary references. Cs$_2$O represents conventional oxide chemistry with little Cs $5p$ participation, whereas high-pressure CsO provides a reference for strong Cs $5p$--O $2p$ bonding. CsO$_2$, described as a Cs(I) superoxide in the previous polyoxide study, provides an oxygen-rich reference with intact O$_2$ units and weak Cs--O bonding~\cite{Sun2021CsPolyoxides}. Calculated Cs $L_3$-edge XANES spectra and powder diffraction patterns provide electronic and structural signatures for experimental identification.

\begin{manuscriptfigure}\centering
\includegraphics[trim=4.000bp 15.400bp 1.228bp 2.000bp,clip,width=\textwidth]{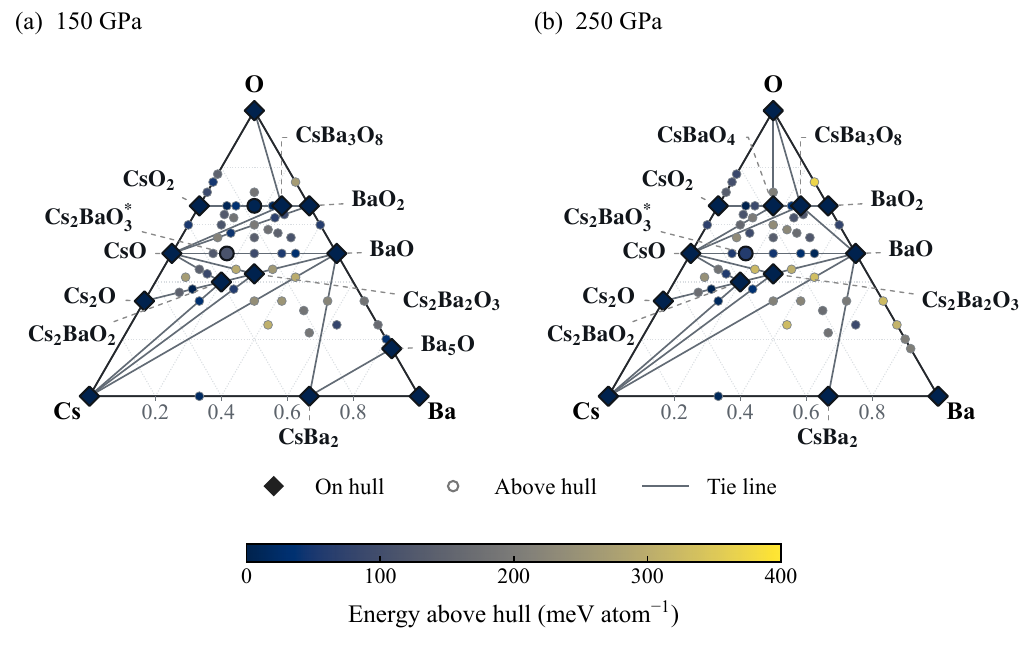}
\caption{\label{fig:hulls}Ternary convex hulls of the Cs--Ba--O system at (a) 150 and (b) 250~GPa. The labeled phases lie on the hull except starred Cs$_2$BaO$_3$, which is included as a metastable candidate.}
\end{manuscriptfigure}

\section{Computational methods}
\label{sec:methods}

Crystal structures were searched using Crystal Structure Analysis by Particle Swarm Optimization (CALYPSO)~\cite{Wang2010PSO} and relaxed at 150 and 250~GPa. Density-functional calculations used the Vienna Ab initio Simulation Package (VASP)~\cite{Kresse1996VASP}, the Perdew--Burke--Ernzerhof functional~\cite{Perdew1996PBE}, and projector augmented-wave (PAW) potentials~\cite{Kresse1999PAW}. The PAW valence configurations were Cs $5s^25p^66s^1$, Ba $5s^25p^66s^2$, and O $2s^22p^4$. Structural optimizations used a 520~eV plane-wave cutoff, a reciprocal-space sampling interval of 0.2~\AA$^{-1}$, and electronic and force convergence thresholds of $10^{-6}$~eV and 0.02~eV~\AA$^{-1}$, respectively. Convex hulls were constructed from the lowest-enthalpy ternary, binary, and elemental structures at each pressure using static-lattice enthalpies, $H=E_{\mathrm{DFT}}+PV$. Zero-point energies and finite-temperature vibrational contributions were not included.

Electronic analyses at 250~GPa included Bader charges~\cite{Bader1994} and electron localization functions~\cite{Becke1990ELF}. Projected densities of states for individual atomic shells, L\"owdin populations, crystal orbital Hamilton populations (COHP), and crystal orbital bond indices (COBI) were calculated with the Local Orbital Basis Suite Towards Electronic-Structure Reconstruction (LOBSTER)~\cite{Nelson2020LOBSTER}. O$_2$ fragment orbitals and their L\"owdin populations were obtained using the linear combination of fragment orbitals (LCFO) method in LOBSTER~\cite{Mueller2024LCFO}. Cs $L_3$-edge x-ray absorption near-edge structure (XANES) spectra were calculated using FEFF9~\cite{Rehr2010FEFF9}. Powder x-ray diffraction (XRD) patterns were generated using Python Materials Genomics (pymatgen)~\cite{Ong2013Pymatgen} with a wavelength of 1.5406~\AA.

\section{Results and discussion}

\subsection{Thermodynamic stability}

The ternary convex hulls of the Cs--Ba--O system at 150 and 250~GPa are shown in Fig.~\ref{fig:hulls}. At 150~GPa, Cs$_2$Ba$_2$O$_3$, Cs$_2$BaO$_2$, and CsBa$_3$O$_8$ are stable against decomposition. All three compounds remain on the hull at 250~GPa, where CsBaO$_4$ also becomes stable. CsBaO$_4$ lies only 4.1~meV atom$^{-1}$ above the hull at 150~GPa and retains the same $Cm$ structure type at both pressures.

Cs$_2$BaO$_3$ remains metastable, but its energy above the hull decreases from 111.7~meV atom$^{-1}$ at 150~GPa to 59.9~meV atom$^{-1}$ at 250~GPa. It shows the strongest Cs $5p$ participation among the ternaries examined and can be related directly to solid oxide precursors. The weaker Cs $5p$ bonding in CsBa$_3$O$_8$ and CsBaO$_4$ occurs in thermodynamically stable phases at 250~GPa. Cs$_2$BaO$_2$ provides a conventional Cs(I) reference for the electronic comparison. Cs$_2$Ba$_2$O$_3$ is also stable at both pressures but is not included in the detailed bonding comparison.

The composition of Cs$_2$BaO$_3$ is obtained directly from a 1:1 mixture of Cs$_2$O and BaO$_2$,
\begin{equation}
\mathrm{Cs_2O + BaO_2 \rightarrow Cs_2BaO_3}.
\label{eq:cs2bao3_formation}
\end{equation}
The reaction enthalpy is $-0.121$~eV per Cs$_2$BaO$_3$ formula at 150~GPa and $-0.594$~eV per formula at 250~GPa. Formation of the ternary compound is therefore favorable relative to these two precursors even though lower-enthalpy products exist on the complete hull. The preferred decomposition is
\begin{equation}
\mathrm{Cs_2BaO_3 \rightarrow 2CsO + BaO},
\label{eq:cs2bao3_decomposition}
\end{equation}
which lowers the enthalpy by 0.670 and 0.360~eV per Cs$_2$BaO$_3$ formula at 150 and 250~GPa, respectively. At 150~GPa, the CsO product retains the peroxide structure. At 250~GPa, CsO adopts the dimer-free $R\overline{3}m$ structure associated with Cs core reactivity~\cite{Sun2021CsPolyoxides}. The corresponding 250~GPa products are shown in Fig.~S1 of the Supplemental Material~\cite{SupplementalMaterial}. CsO and BaO are the lowest-enthalpy products at this composition, while Cs$_2$BaO$_3$ remains a metastable ternary candidate with strong Cs $5p$ participation.

The stable ternaries also have negative reaction enthalpies relative to the binary oxide mixtures listed in Table~S1. The CsO$_2$--BaO$_2$ comparisons evaluate relative stability; they do not establish practical precursor routes for CsBa$_3$O$_8$ or CsBaO$_4$.

\begin{manuscriptfigure}
\centering
\includegraphics[trim=8.000bp 4.134bp 13.543bp 14.000bp,clip,width=\textwidth]{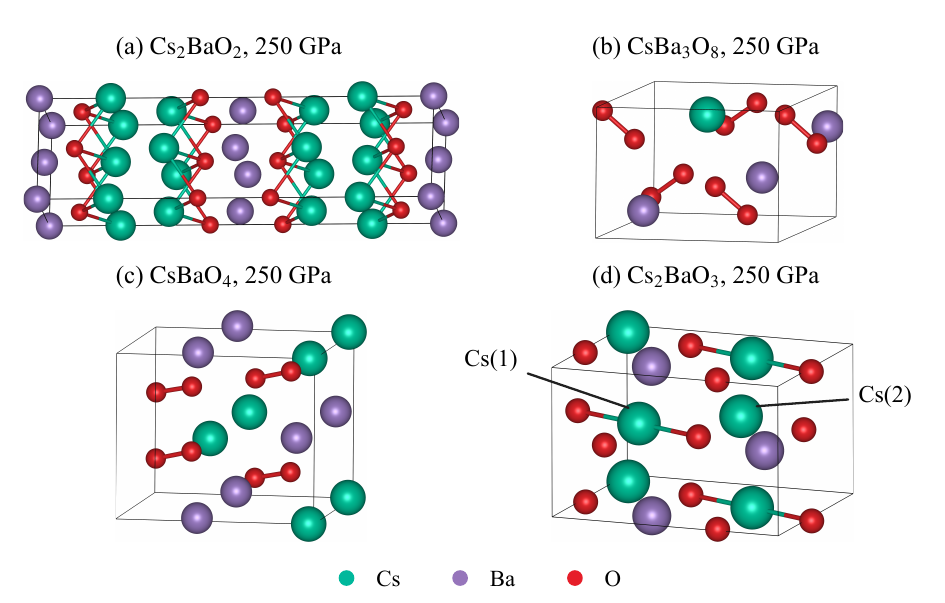}
\caption{\label{fig:structures}Crystal structures of (a) Cs$_2$BaO$_2$, (b) CsBa$_3$O$_8$, (c) CsBaO$_4$, and (d) Cs$_2$BaO$_3$ at 250~GPa.}
\end{manuscriptfigure}

\par\noindent\begin{minipage}{\columnwidth}
\setlength{\parindent}{1em}
\vspace*{14pt}
\subsection{Crystal structures}

Figure~\ref{fig:structures} compares the structures at 250~GPa, the pressure used for the electronic analysis. Lattice parameters for the ternaries and binary references at both pressures are given in Table~S2, and the local Cs geometries in Table~S3.
\end{minipage}\par

Cs$_2$BaO$_2$ adopts the tetragonal $I4/mmm$ structure at both pressures and contains no O$_2$ units. In contrast, $P1$ CsBa$_3$O$_8$ and $Cm$ CsBaO$_4$ contain O$_2$ units with shortest O--O distances of 1.329 and 1.311~\AA, respectively. The $Cm$ Cs$_2$BaO$_3$ structure is also dimer free, with a shortest O--O separation of 2.557~\AA.

Cs$_2$BaO$_3$ contains two inequivalent Cs sites, labeled Cs(1) and Cs(2) in Fig.~\ref{fig:structures}(d). Both form nearly linear O--Cs--O units, with shorter Cs--O contacts at Cs(1). This arrangement resembles core-reactive CsO, whose two Cs sites likewise differ in their nearest Cs--O distances~\cite{Sun2021CsPolyoxides}. The individual bond lengths and angles are listed in Table~S3.

\subsection{Electronic structure and Cs--O bonding}

The projected density of states at 250~GPa shows how Cs $5p$ participation changes across the four compounds (Fig.~\ref{fig:pdos}). In Cs$_2$BaO$_2$, the Cs $5p$ states are almost entirely occupied and lie below the Fermi level, with negligible weight in the unoccupied states.

In the stable CsBa$_3$O$_8$ and CsBaO$_4$ phases, Cs $5p$ and O $2p$ states overlap throughout the upper valence region, and a small Cs $5p$ contribution appears immediately above $E_{\mathrm F}$. Cs$_2$BaO$_3$ shows a much larger unoccupied feature near 1~eV with both Cs $5p$ and O $2p$ character. This feature resembles that found in core-reactive CsO~\cite{Sun2021CsPolyoxides} and is consistent with stronger involvement of the Cs $5p$ shell in bonding.

Table~\ref{tab:cs-charges} compares the Cs charges and $5p$ populations in the ternaries and the three binary references at 250~GPa. Bader analysis assigns a partial charge to each atom from its surrounding electron density~\cite{Bader1994}. These charges describe electron redistribution rather than formal oxidation states.

L\"owdin analysis complements this comparison by giving the total atomic charge and the populations of individual shells~\cite{Ertural2019Charges}. We define the $5p$ orbital charge as $q_{5p}=6-n_{5p}$, in units of $e$, where $n_{5p}$ is the Cs $5p$ population. It measures electron depletion relative to a filled $5p^6$ shell. The shell populations refer to the orthogonalized local basis and are interpreted relative to the binary references together with the Cs--O bonding analysis. The Cs $5s$, $5p$, and $6s$ populations are listed in Table~S4.
\begin{manuscriptfigure}
\centering
\includegraphics[width=0.88\textwidth]{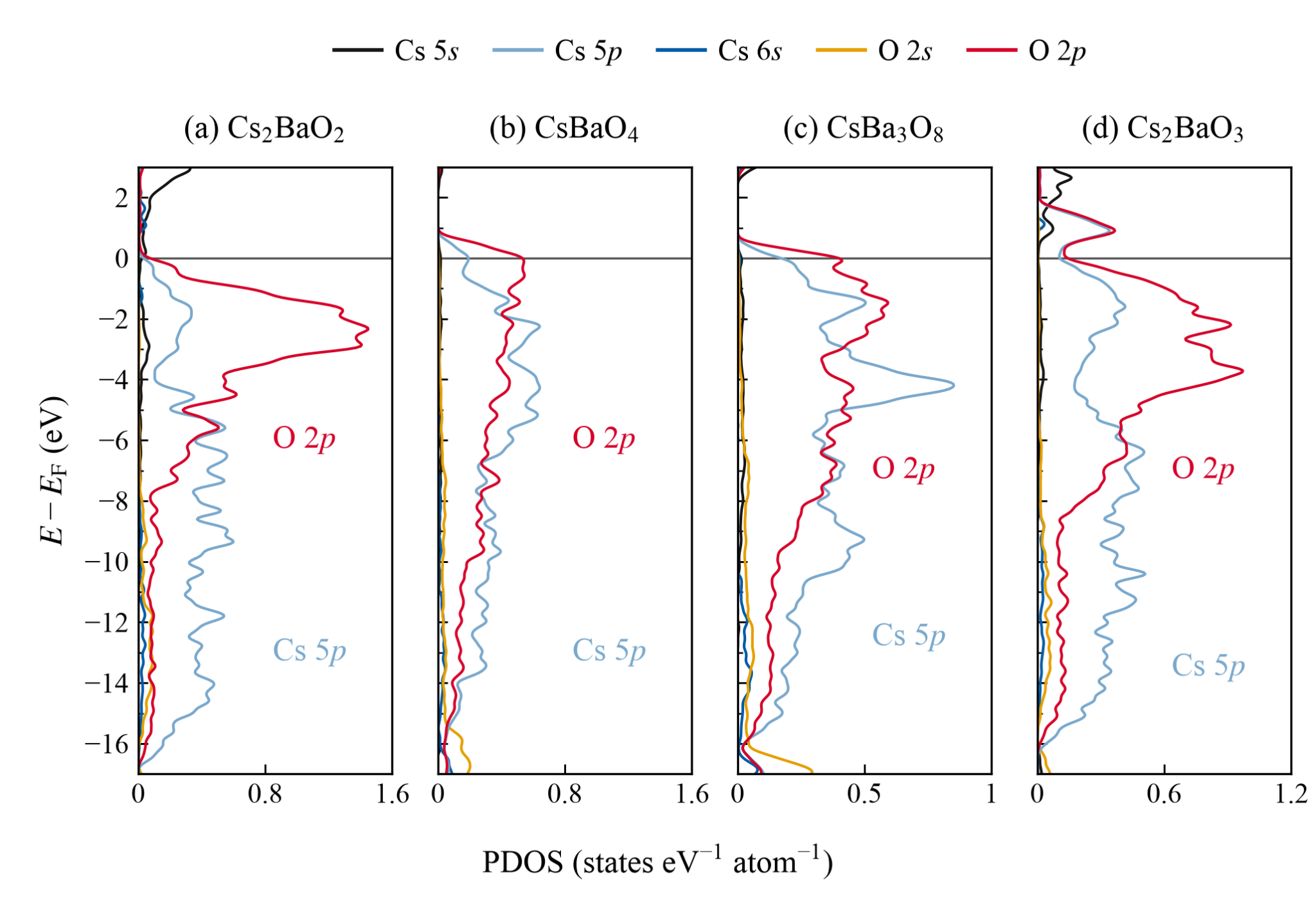}
\caption{\label{fig:pdos}Projected densities of states of (a) Cs$_2$BaO$_2$, (b) CsBaO$_4$, (c) CsBa$_3$O$_8$, and (d) Cs$_2$BaO$_3$ at 250~GPa.}
\end{manuscriptfigure}

Cs$_2$BaO$_2$ closely resembles Cs$_2$O in both its partial charge and $5p$ population. CsBa$_3$O$_8$ has nearly the same $q_{5p}$ as CsO$_2$, 0.211 and 0.218~$e$, respectively, while CsBaO$_4$ is modestly higher at 0.255~$e$. The Bader charges of both oxygen-rich ternaries are also close to that of CsO$_2$, although their total L\"owdin charges are lower (Table~\ref{tab:cs-charges}).

Cs$_2$BaO$_3$ shows the largest $q_{5p}$ values among the ternaries. Cs(1) has a Bader charge of $+1.127\,e$ and $q_{5p}=0.519\,e$, compared with $+1.004\,e$ and $0.384\,e$ for Cs(2). The larger charges at the shorter-bonded Cs(1) site follow the site ordering in core-reactive CsO~\cite{Sun2021CsPolyoxides}.

The Ba Bader charges are comparable to that in BaO, consistent with conventional Ba(II) character in the ternaries (Table~S5).

\begin{manuscripttable}
\normalsize
\caption{Cs Bader charges ($q_{\mathrm B}$), L\"owdin charges ($q_{\mathrm L}$), and $5p$ orbital charges ($q_{5p}$) at 250~GPa.}
\label{tab:cs-charges}
\centering
\renewcommand{\arraystretch}{1.35}
\begin{ruledtabular}
\begin{tabular}{llccc}
Compound & Site & $q_{\mathrm{B}}$ ($e$) & $q_{\mathrm{L}}$ ($e$) & $q_{5p}$ ($e$) \\
\hline
Cs$_2$BaO$_2$ & Cs & 0.654 & 0.834 & 0.085 \\
CsBa$_3$O$_8$ & Cs & 0.903 & 0.883 & 0.211 \\
CsBaO$_4$ & Cs & 0.939 & 0.974 & 0.255 \\
Cs$_2$BaO$_3$ & Cs(1) & 1.127 & 1.327 & 0.519 \\
 & Cs(2) & 1.004 & 1.191 & 0.384 \\
\hline
Cs$_2$O & Cs & 0.646 & 0.867 & 0.097 \\
CsO$_2$ & Cs & 0.912 & 1.037 & 0.218 \\
CsO & Cs(1) & 1.360 & 1.633 & 0.774 \\
 & Cs(2) & 0.846 & 1.089 & 0.243 \\
\end{tabular}
\end{ruledtabular}
\end{manuscripttable}

\begin{manuscriptfigure}
\centering
\includegraphics[trim=9.000bp 12.496bp 10.370bp 0.000bp,clip,width=0.82\textwidth]{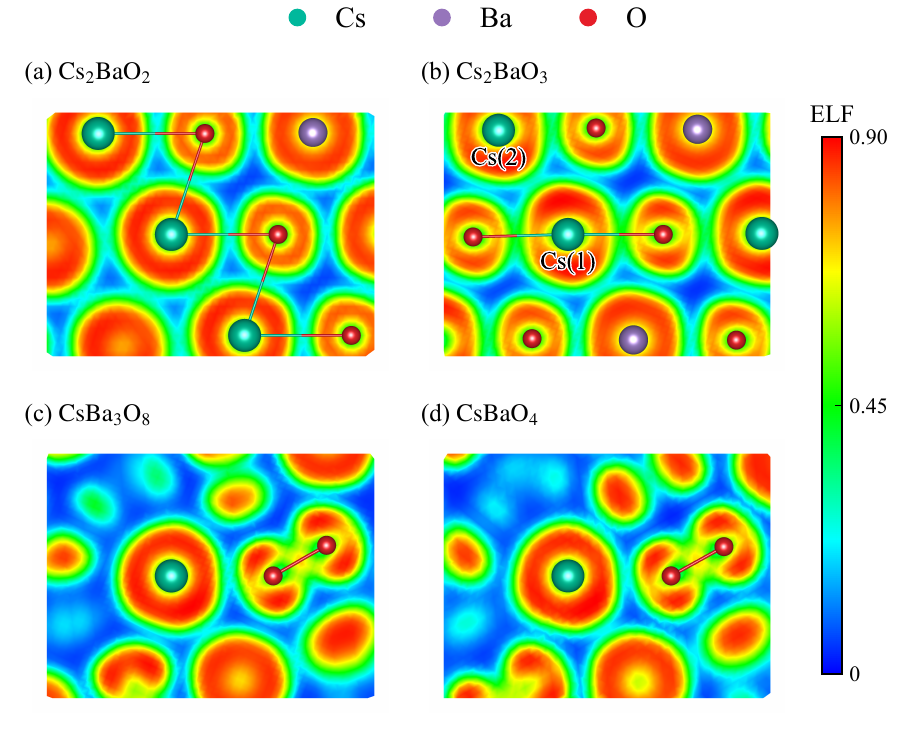}
\caption{\label{fig:elf}ELF sections through selected Cs--O contacts in (a) Cs$_2$BaO$_2$, (b) Cs$_2$BaO$_3$, (c) CsBa$_3$O$_8$, and (d) CsBaO$_4$ at 250~GPa. Panel (b) is the $(010)$ section.}
\end{manuscriptfigure}

The electron localization function (ELF) further distinguishes the Cs--O interactions (Fig.~\ref{fig:elf}). In Cs$_2$BaO$_2$, regions of high ELF surround the atoms and are separated along the Cs--O contacts. CsBa$_3$O$_8$ and CsBaO$_4$ retain pronounced localization within their O$_2$ units, while localization between Cs and O remains much weaker.

In Cs$_2$BaO$_3$, enhanced localization develops along the nearly linear O--Cs--O units [Fig.~\ref{fig:elf}(b)]. The greater interatomic localization along Cs(1)--O is consistent with the shorter Cs--O distances and larger $q_{5p}$ at this site.

Crystal orbital Hamilton population (COHP) and crystal orbital bond index (COBI) analyses quantify the contribution of the Cs $5p$ shell to Cs--O bonding~\cite{Dronskowski1993COHP,Muller2021COBI}. Integration over occupied states gives ICOHP, which provides a measure of bond strength, and ICOBI, a dimensionless index related to covalent bond order. More negative ICOHP values indicate stronger bonding interactions, while larger ICOBI values indicate greater covalent bonding. Figure~\ref{fig:cohp-main} compares selected Cs--O $-$COHP curves in all four ternaries, and Table~\ref{tab:cs-orbital-bonding} summarizes the integrated Cs $5p$--O $2p$ contributions. The complete COHP/COBI comparison and contact data are provided in Fig.~S2 and Table~S6.

Cs$_2$BaO$_2$ closely resembles Cs$_2$O, with negligible Cs $5p$--O $2p$ covalent bonding. CsBa$_3$O$_8$, CsBaO$_4$, and CsO$_2$ have comparable $5p$--$2p$ ICOBI values near 0.028, while the corresponding ICOHP contributions are more negative in the ternaries (Table~\ref{tab:cs-orbital-bonding}). Together with their reduced $5p$ populations, these finite bonding contributions support mild Cs core participation in all three compounds.

\begin{manuscriptfigure}
\centering
\includegraphics[width=0.88\textwidth]{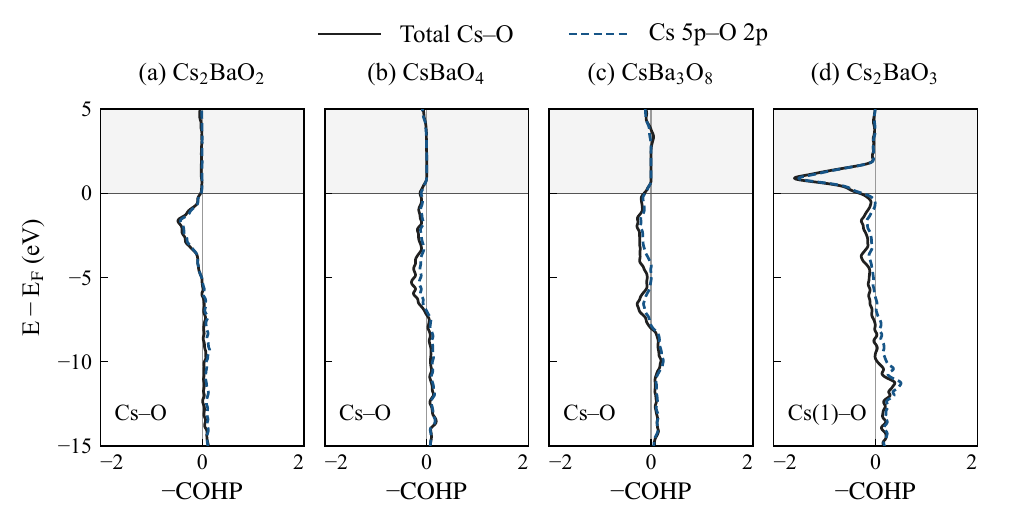}
\caption{\label{fig:cohp-main}$-$COHP for selected Cs--O contacts in (a) Cs$_2$BaO$_2$, (b) CsBaO$_4$, (c) CsBa$_3$O$_8$, and (d) Cs$_2$BaO$_3$ at 250~GPa. Panel (d) shows Cs(1)--O. Positive and negative values denote bonding and antibonding contributions, respectively.}
\end{manuscriptfigure}

\begin{manuscriptfigure}
\centering
\includegraphics[trim=19.000bp 8.811bp 0.000bp 13.000bp,clip,width=0.72\textwidth]{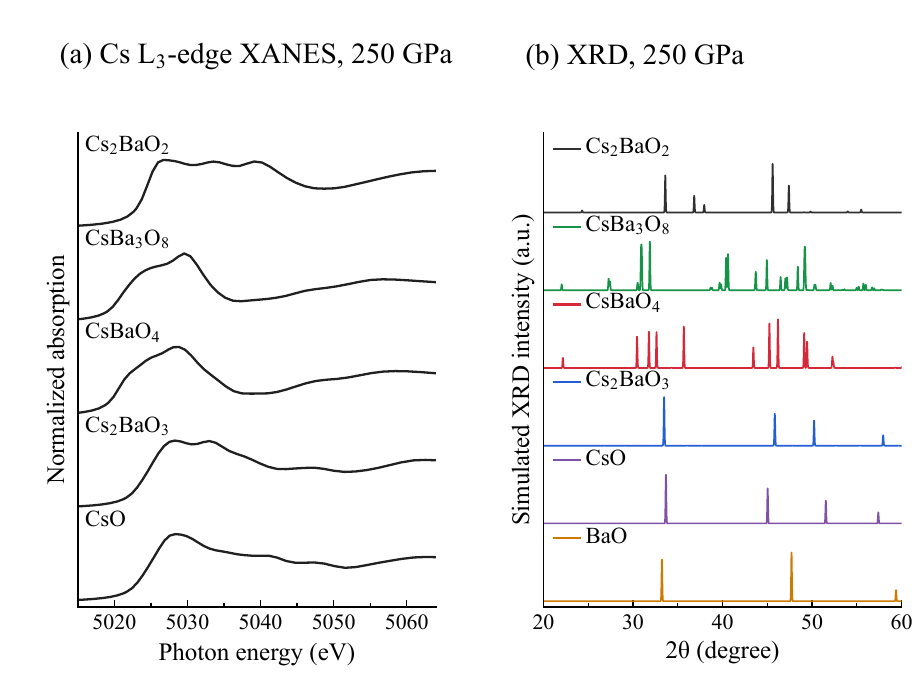}
\caption{\label{fig:xanes-xrd}Calculated (a) Cs $L_3$-edge XANES spectra and (b) powder XRD patterns at 250~GPa. The XRD wavelength is 1.5406~\AA.}
\end{manuscriptfigure}

\begin{manuscripttable}
\normalsize
\caption{Cs $5p$--O $2p$ contributions to ICOHP and ICOBI for selected Cs--O contacts at 250~GPa.}
\label{tab:cs-orbital-bonding}
\centering
\renewcommand{\arraystretch}{1.35}
\begin{ruledtabular}
\begin{tabular}{llrr}
Compound & Site & \shortstack{ICOHP\\(eV/pair)} & ICOBI \\
\hline
Cs$_2$BaO$_2$ & Cs & $-0.035$ & 0.0004 \\
CsBa$_3$O$_8$ & Cs & $-0.428$ & 0.0278 \\
CsBaO$_4$ & Cs & $-0.396$ & 0.0264 \\
Cs$_2$BaO$_3$ & Cs(1) & $-1.557$ & 0.1999 \\
 & Cs(2) & $-1.120$ & 0.1296 \\
\hline
Cs$_2$O & Cs & $-0.059$ & 0.0008 \\
CsO$_2$ & Cs & $-0.346$ & 0.0280 \\
CsO & Cs(1) & $-2.533$ & 0.3358 \\
 & Cs(2) & $-0.619$ & 0.0463
\end{tabular}
\end{ruledtabular}
\end{manuscripttable}

Cs$_2$BaO$_3$ shows substantially stronger Cs $5p$--O $2p$ bonding at both inequivalent Cs sites. For the selected Cs(1)--O contact, the $5p$--$2p$ interaction dominates the occupied bonding states and the unoccupied antibonding feature immediately above $E_{\mathrm F}$ [Fig.~\ref{fig:cohp-main}(d)]. This channel accounts for approximately 90\% of the total ICOHP magnitude and 94\% of the ICOBI.

Cs(1) is the more strongly bonded site in both Cs$_2$BaO$_3$ and core-reactive CsO~\cite{Sun2021CsPolyoxides}, but the contrast between sites is smaller in the ternary. Its Cs(1)--O interaction is weaker than in CsO, whereas Cs(2)--O is stronger, as shown by both ICOHP and ICOBI (Table~\ref{tab:cs-orbital-bonding}).

The character of the retained O$_2$ units can be assessed from their antibonding $\pi^*$ populations. The two $\pi^*$ orbitals contain three electrons in ideal superoxide and four in peroxide~\cite{Radjenovic2019Dioxides}. The calculated mean populations increase from 3.08 electrons in CsO$_2$ to 3.46 in CsBaO$_4$ and 3.66 in CsBa$_3$O$_8$ (Table~S7), indicating greater peroxide-like character in the ternaries. Their more negative dimer charges (Table~S5) and longer, weaker O--O bonds (Table~S6) follow the same trend. Conversely, Cs$_2$BaO$_2$ has no O$_2$ units but little Cs $5p$ bonding, so dimer absence alone does not establish core activation.

The stable oxygen-rich ternaries combine weak Cs $5p$--O $2p$ bonding with O$_2$ units that are more peroxide-like than those in CsO$_2$. Cs$_2$BaO$_3$ exhibits substantially stronger Cs $5p$--O $2p$ bonding but remains metastable against decomposition into CsO and BaO.

\subsection{Prospects for synthesis and experimental identification}

The ternary phase diagram identifies CsBa$_3$O$_8$ and CsBaO$_4$ as stable targets for observing Cs $5p$ participation at 250~GPa. At the Cs$_2$O--BaO$_2$ composition, CsO and BaO are the lowest-enthalpy products, while Cs$_2$BaO$_3$ is a competing ternary candidate with strong Cs $5p$ participation. Laser-heated diamond anvil cell (DAC) synthesis of the predicted high-pressure MgO$_2$ phase provides a precedent for exploring unusual oxide chemistry through compression and heating~\cite{Lobanov2015MgPeroxide}. The required synthesis conditions remain to be established experimentally.

The Cs $L_3$-edge XANES profiles of Cs$_2$BaO$_3$ and core-reactive CsO are broadly similar, consistent with their related Cs--O environments [Fig.~\ref{fig:xanes-xrd}(a)]. Their near-edge shapes differ, and the profiles also depend on local coordination and covalency~\cite{Bauters2020XANES,Biffar2025CsXANES}.

The diffraction patterns [Fig.~\ref{fig:xanes-xrd}(b)] provide structural references for distinguishing Cs$_2$BaO$_3$ from CsO and BaO, with XANES providing complementary information about the Cs environment.

\section{Conclusion}

First-principles calculations identify thermodynamically stable Cs--Ba--O compounds with weak Cs $5p$ participation and a metastable Cs$_2$BaO$_3$ phase with substantially stronger covalent Cs--O bonding. In the stable oxygen-rich ternaries, Cs core participation coexists with intact O$_2$ units whose greater antibonding occupation gives them increased peroxide-like character relative to CsO$_2$.

The Cs$_2$O--BaO$_2$ mixture offers an alternative solid starting composition for investigating Cs core reactivity. Formation of Cs$_2$BaO$_3$ is favorable relative to the starting oxides, although CsO and BaO are the lowest-enthalpy products. Since CsO is core reactive at 250~GPa, the composition remains relevant even if the ternary does not form. The diffraction patterns provide structural references for phase identification, while XANES offers a complementary comparison of the Cs environments.

\begin{acknowledgments}
This work was supported by the U.S. Department of Defense Research and Education Program for Historically Black Colleges and Universities and Minority-Serving Institutions Basic Research Funding (No. W911NF2310232); the National Science Foundation (DMR-1848141 and OAC-2117956); the Camille and Henry Dreyfus Foundation; and California State University Research, Scholarship and Creative Activity awards.

During preparation of this manuscript, generative AI tools, including ChatGPT (OpenAI; GPT-6 Astra Pro), were used solely to assist with language editing, organization, and formatting. The authors reviewed and edited all AI-assisted content and accept full responsibility for the manuscript. No AI tool was used to generate primary scientific data, perform the reported analyses, or draw the scientific conclusions.
\end{acknowledgments}

\bibliography{references}

\end{document}